\documentstyle[12pt]{article}
\def\PRL #1 #2 #3{{\sl Phys. Rev. Lett.} {\bf#1} (#2) #3}
\def\NPB #1 #2 #3{{\sl Nucl. Phys.} {\bf B #1} (#2) #3}
\def\NPBFS #1 #2 #3 #4 {{\sl Nucl. Phys.} {\bf B #2} [FS#1] (#3) #4}
\def\CMP #1 #2 #3 {{\sl Commun.  Math. Phys.} {\bf #1} (#2) #3}
\def\PRD #1 #2 #3 {{\sl Phys.  Rev.} {\bf D #1} (#2) #3}
\def\PLA #1 #2 #3 {{\sl Phys. Lett.} {\bf #1A} (#2) #3}
\def\PLB #1 #2 #3 {{\sl Phys.  Lett.} {\bf B #1} (#2) #3}
\def\JMP #1 #2 #3 {{\sl J. Math.  Phys.} {\bf #1} (#2) #3}
\def\PTP #1 #2 #3 {{\sl Prog. Theor. Phys.} {\bf #1} (#2) #3}
\def\SPTP #1 #2 #3 {{\sl Suppl.  Prog.  Theor. Phys.} {\bf #1} (#2) #3}
\def\AoP #1 #2 #3 {{\sl Ann. of Phys.} {\bf #1} (#2) #3}
\def\PNAS #1 #2 #3 {{\sl Proc.  Natl. Acad. Sci.  USA} {\bf #1} (#2) #3}
\def\RMP #1 #2 #3 {{\sl Rev. Mod.  Phys.} {\bf #1} (#2) #3}
\def\PR #1 #2 #3 {{\sl Phys. Reports} {\bf #1} (#2) #3}
\def\AoM #1 #2 #3 {{\sl Ann. of Math.} {\bf #1} (#2) #3}
\def\UMN #1 #2 #3 {{\sl Usp.  Mat.  Nauk} {\bf #1} (#2) #3}
\def\FAP #1 #2 #3 {{\sl Funkt. Anal.  Prilozheniya} {\bf #1} (#2) #3}
\def\FAaIA #1 #2 #3 {{\sl Functional Analysis and Its Application}
{\bf #1} (#2) #3}
\def\BAMS #1 #2 #3 {{\sl Bull. Am. Math.  Soc.} {\bf #1} (#2) #3}
\def\TAMS #1 #2 #3 {{\sl Trans.  Am. Math. Soc.} {\bf #1} (#2) #3}
\def\InvM #1 #2 #3 {{\sl Invent. Math.} {\bf #1} (#2) #3}
\def\LMP #1 #2 #3 {{\sl Letters in Math.  Phys.} {\bf #1} (#2) #3}
\def\IJMPA #1 #2 #3 {{\sl Int. J. Mod. Phys.} {\bf A #1} (#2) #3}
\def\AdM #1 #2 #3 {{\sl Advances in Math.} {\bf #1} (#2) #3}
\def\RMaP #1 #2 #3 {{\sl Reports on Math. Phys.} {\bf #1} (#2) #3}
\def\IJM #1 #2 #3 {{\sl Ill. J.  Math.} {\bf #1} (#2) #3}
\def\APP #1 #2 #3 {{\sl Acta Phys.  Polon.} {\bf #1} (#2) #3}
\def\TMP #1 #2 #3 {{\sl Theor.  Mat.  Phys.} {\bf #1} (#2) #3}
\def\JPA #1 #2 #3 {{\sl J. Physics} {\bf A#1} (#2) #3}
\def\JSM #1 #2 #3 {{\sl J. Soviet Math.} {\bf #1} (#2) #3}
\def\MPLA #1 #2 #3 {{\sl Mod.  Phys. Lett.} {\bf A #1} (#2) #3}
\def\JETP #1 #2 #3 {{\sl Sov. Phys. JETP} {\bf #1} (#2) #3}
\def\JETPL #1 #2 #3 {{\sl Sov. Phys.  JETP Lett.} {\bf #1} (#2) #3}
\def\PHSA #1 #2 #3 {{\sl Physica} {\bf A #1} (#2) #3}
\def\CQG #1 #2 #3 {{\sl Class. Quantum Grav.} {\bf #1} (#2) #3}
\def\SJNP #1 #2 #3 {{\sl Sov. J. Nucl. Phys.  (Yadern.Fiz.)} {\bf #1}
(#2) #3}
\def\a{\alpha}\def\b{\beta}\def\g{\gamma}\def\d{\delta}\def\e{\epsilon}

 \def\k{\kappa}\def\s{\sigma}
\def\G{\Gamma}

\newcommand{\p}[1]{(\ref{#1})}
\begin{document}
\renewcommand{\thefootnote}{\fnsymbol{footnote}}
\begin{flushright}
{\bf DFPD/97/04 \\
hep-th/9701xxx} \\
1997, January \\
\end{flushright}

\vspace{1.5cm}

\def\Ai{\hbox{\hbox{${\cal A}$}}\kern-1.9mm{\hbox{${/}$}}}
\def\Vi{\hbox{\hbox{${\cal V}$}}\kern-1.9mm{\hbox{${/}$}}}
\def\Di{\hbox{\hbox{${\cal D}$}}\kern-1.9mm{\hbox{${/}$}}}
\def\lam{\hbox{\hbox{${\lambda}$}}\kern-1.6mm{\hbox{${/}$}}}
\def\D{\hbox{\hbox{${D}$}}\kern-1.9mm{\hbox{${/}$}}}
\def\A{\hbox{\hbox{${A}$}}\kern-1.8mm{\hbox{${/}$}}}
\def\V{\hbox{\hbox{${V}$}}\kern-1.9mm{\hbox{${/}$}}}
\def\parz{\hbox{\hbox{${\partial}$}}\kern-1.7mm{\hbox{${/}$}}}
\def\B{\hbox{\hbox{${B}$}}\kern-1.7mm{\hbox{${/}$}}}
\def\R{\hbox{\hbox{${R}$}}\kern-1.7mm{\hbox{${/}$}}}
\def\si{\hbox{\hbox{${\xi}$}}\kern-1.7mm{\hbox{${/}$}}}

\null
\vskip 0.3truecm

\begin{center}

{\Large \bf GENERALIZED ACTION PRINCIPLE and}

{\Large \bf SUPERFIELD EQUATIONS OF MOTION for}

{\Large \bf D=10  D--p--BRANES}

\vspace{1.0cm}

{\bf Igor A. Bandos, Dmitri P. Sorokin }
\\
{\it Institute for Theoretical Physics, \\
NSC Kharkov Institute of Physics and Technology \\
Akademicheskaya str. 1, 310108, Kharkov \\
Ukraine} \\
{\bf e-mail: kfti@rocket.kharkov.ua}

\vspace{0.7cm}
\vspace{0.3cm}

{\bf and}

\vspace{0.7cm}
 {\bf Mario Tonin}
\\
\vspace{0.2cm}
{\it Universit\`a Degli Studi Di Padova \\
Dipartimento Di Fisica ``Galileo Galilei''\\
and INFN, Sezione Di Padova \\
Via F. Marzolo, 8, 35131 Padova \\ Italia}\\
{\bf e-mail tonin@padova.infn.it}

\end{center}

\vskip 1truecm
\centerline{\bf ABSTRACT}

\vskip 0.5truecm
The action for the $D=10$ type II Dirichlet super--p--branes, which has
been obtained recently, is reconstructed in a more geometrical form
involving Lorentz harmonic variables. This new (Lorentz harmonic)
formulation possesses $\kappa$--symmetry in an irreducible
form and is used as a basis for applying a generalized action principle
that provides the superfield equations of motion and clarifies the
geometrical nature of the $\kappa$--symmetry of these models. The case of
a Dirichlet super--3--brane is considered in detail.

\newpage
\renewcommand{\thefootnote}{\arabic{footnote}}
\section{Introduction}

\bigskip

Recently  $\kappa$--invariant actions for the  $D=10$ $type~IIB$
Dirichlet super--3--brane \cite{c1} and then for all $D=10~type~II$
Dirichlet super--p--branes  (D--p--branes) \cite{c2}--\cite{bt}
were obtained. These actions consist of the sum of a Dirac--Born--Infeld
(DBI) action and a Wess--Zumino (WZ) term. The role of fermionic
$\kappa$--symmetry in these models is to reduce half the number of
components of the target space spinors. Therefore the $\kappa$--symmetry
variation of the spinors involves a projector given in terms of a
traceless matrix $\bar\Gamma$ that squares to unity. A remarkable
property of the models \cite{c1}--\cite{bt} is that the $\kappa$--symmetry
variation of the DBI action can be written as the integral of a
$(p+1)$--form and hence can compensate the variation of the WZ term.  This
property looks miraculous from the point of view of the papers
\cite{c1}--\cite{bt}.

In the present paper, by use of Lorentz harmonics
\cite{sok} -- \cite{bsv}
as  auxiliary
variables, we  rewrite the full action of these models as the integral
of a Lagrangian $(p+1)$--form over a $d=p+1$ dimensional worldvolume (see
\cite{bpstv,bsv} for superstrings and $type~I$ super--$p$--branes) and
 verify its $\kappa$--invariance. In this way the remarkable property
mentioned above appears quite naturally and its geometrical nature is
clarified.

The formulation we propose is a generalization to the case of
super--$D$--$p$--branes of a geometrical twistor--like approach to
describing supersymmetric extended objects developed in
\cite{bpstv,bsv,hs1,hs2}.

Since our action is written in terms of differential forms without any use
of  Hodge operation it extends in a straightforward way to a group
manifold (or generalized) action
\cite{rheo,bsv}
that describes the embedding of the brane
superworldvolume in the target superspace \cite{bpstv,hs1,hs2}.

This is done simply by replacing  the purely bosonic worldvolume with an
arbitrary (p+1)--dimensional surface in the whole superworldvolume and
regarding the coordinate functions and supervielbein components as
worldvolume superfields restricted to this arbitrary  surface. Then a
generalized action principle produces the superspace field equations
typical to the twistor--like formulations and, as in the case of
superparticles, superstrings and type I super--p--branes, gives the
geometrical meaning of the $\kappa$--symmetry in these models
as a manifestation of worldvolume superdiffeomorphisms
\cite{stv}--\cite{bers94}. Thus our approach provides a bridge between
the formulation of refs.  \cite{c1}--\cite{bt} and the superspace approach
of ref.  \cite{hs1,hs2}.

For simplicity the details will be worked out only for a
super--$D$--$3$--brane but the generalization to other super-D-branes
is straightforward.

The paper is organized as follows.
In Section {\bf 2} we fix our
notation and introduce Lorentz harmonics \cite{sok}--\cite{bpstv}.
In Section {\bf 3} we describe the DBI action for
super--$D$--$p$--branes \cite{c1,c2}.
 In Section {\bf 4} we propose the new formulation for
 Dirichlet super--p--branes which involves Lorentz--harmonic variables.
 Section {\bf 5} is devoted to the proof of the $\kappa$--invariance
 of the Lorentz--harmonic action.
 In Section {\bf 6} we demonstrate that by use of the Lorentz
 harmonics $\kappa$--symmetry can be rewritten in an irreducible form
 and present the set of
 16 covariant parameters of irreducible $\kappa$--symmetry for the
 $type~IIB$ super--3--brane.

 In Section {\bf 7} we use the Lorentz--harmonic formulation for
 the construction of the generalized action for super--D-p--branes in
 $D=10$ type $II$ supergravity background and obtain superfield equations
 of motion for these objects. Equations of motion for a super--3--brane in
 flat $D=10$ $type ~IIB$ superspace are analyzed in more detail in
 Section {\bf 8}.

\bigskip

\section{Notation and conventions}

\bigskip

First of all let us describe our notation. We shall use underlined
indices for target (super)space and not underlined ones for world
(super)surface. Latin and Greek letters denote vector and spinor indices
respectively and letters from the beginning or the middle of the alphabet
refer respectively to tangent space or curved spaces.
The hat and/or tilde over the index denote its reducible structure
with respect to the Lorentz group, namely, a $32$--dimensional $type~II$
spinor index \cite{c1,c2} and a composite spinor index of
$SO(1,p)\times SO(9-p)$ respectively (see below).

The supervielbeins of $D=10$ $N=2$  target
superspace are denoted as
$$
E^{\underline{A}} = d Z^{\underline{M}} E^{\underline{A}}_{\underline{M}}
(Z) \equiv (E^{\underline{a}}, \hat{E}^{\hat{\underline{\alpha}}}),
$$
where $Z^{\underline{M}} \equiv (x^{\underline{m}},
\theta^{\hat{\underline{\mu}}})~~$
(${\hat{\underline{m}}}=0,...,9,~~{\hat{\underline{\mu}}}=1,...,32$)
are the local coordinates of the $type~ II$ superbrane,
$\underline{a} = 0, 1, ...  9$ is a $D=10$ vector index, and
$\hat{\underline{\alpha}}$ is a
32--valued Majorana index of $SO (1,9)$ for II A models or a composite
spinor index of $SO (1,9) \times SO (2)$ for II B models.

\bigskip

The spinor index is reducible with respect to the $SO(1,9)$
Lorentz group and can be decomposed into two $16$--valued indices.
In the  $type~IIA$ case this decomposition
corresponds to the splitting
of a $D=10$ Majorana spinor into two Majorana--Weyl
spinors  of opposite chiralities
\begin{equation}\label{0.0}
\hat{E}^{\hat{\underline{\alpha}}} =
(E^{\underline{\alpha}1} , E^{2}_{\underline{\alpha}}), \qquad
\underline{\alpha}=1,...,16 \end{equation}
For  $type~IIB$ case
it is convenient to
use splitting
\begin{equation}\label{0.2}
\hat{E}^{\hat{\underline{\alpha}}} =
(E^{\underline{\alpha}} ,
\bar{E}^{\underline{\alpha}}) \equiv
(E^{\underline{\alpha}1} + i E^{\underline{\alpha}2},
E^{\underline{\alpha}1} - i E^{\underline{\alpha}2}),
\qquad \underline{\alpha}=1,...,16
\end{equation}
which possesses a complex structure  inherent to $IIB$ superspace.
(Note that in refs. \cite{c1,c2} real splitting of the spinor components
\begin{equation}\label{0.1}
\hat{E}^{\hat{\underline{\alpha}}} =
(E^{\underline{\alpha}1} , E^{{\underline{\alpha}}2}), \qquad
{\underline{\alpha}}=1,...,16
\end{equation}
was implied).

The representation of gamma matrices corresponding to the decomposition
\p{0.2} is
\begin{equation}\label{0.3}
(\Gamma^{\underline{a}})_{\hat{\underline{\a}}\hat{\underline{\b}}} =
\s^{\underline{a}}_{\underline{\a}\underline{\b}}
\otimes K
\end{equation}
where the matrix $K$ belongs  to a set $(K, J, I)$ introduced in
\cite{c1,c2}. These matrices have the form
\begin{equation}\label{0.4}
K =
\left(\matrix{ 0 & 1 \cr 1 & 0  \cr } \right) , \qquad
J = i \left(\matrix{ 0 & 1 \cr - 1 & 0  \cr } \right) , \qquad
I = - i \left(\matrix{ 1 & 0 \cr 0 & -1  \cr } \right) , \qquad
\end{equation}
in the representation \p{0.2}, which corresponds to
\begin{equation}\label{0.5}
K = \left(\matrix{ 1 & 0 \cr 0 & -1  \cr } \right) , \qquad
J = \left(\matrix{ 0 & 1 \cr 1 & 0  \cr } \right) , \qquad
I = \left(\matrix{ 0 & 1 \cr -1 & 0  \cr } \right) , \qquad
\end{equation}
in the real representation \p{0.1}.
In \p{0.3}
$
\s^{\underline{a}}_{\underline{\a}\underline{\b}}
$
are $16\times 16$ Majorana--Weyl $\gamma$--matrices whose
$SO(1,p)\otimes SO(9-p)$ invariant representation can be chosen in
a form which reflects a complex structure inherent in the worldvolume
 superspace of the $D$--$p$--brane.
 For instance, for $p=3$ it is convenient to choose the following
 representation
 $$
\s^{\underline{a}}_{\underline{\a}\underline{\b}} =
(
\s^{a}_{\underline{\a}\underline{\b}} ,
\s^{i}_{\underline{\a}\underline{\b}} ) , \qquad {a}=0,...,3
\qquad i = 1,...,6 \qquad {\underline{a}} = 0,...,9
 $$
 \begin{equation}\label{0.51}
\s^{a}_{\underline{\a}\underline{\b}}
=
\left(\matrix{ 0 & \s^a_{\a \dot{\b}} \d_{~p}^q \cr
    \s^a_{\b\dot{\a}} \d^{~p}_{q} & 0  \cr } \right) ,
\qquad
(\tilde{\s}^{a})^{\underline{\a}\underline{\b}}
=
\left(\matrix{ 0 & (\tilde{\s}^a)^{\dot{\b}\a} \d^{~p}_q \cr
    (\tilde{\s}^a)^{\dot{\a}\b} \d_{~p}^{q} & 0  \cr } \right) ,
\qquad
 \end{equation}
 $$
 \a,\b =1,2 \qquad \dot{\a}, \dot{\b} =1,2 \qquad q, p = 1,...,4
$$
$$
\s^{i}_{\underline{\a}\underline{\b}}
=
\left(\matrix{ \e_{\a\b}(\tilde{\g}^i)^{qp} & 0 \cr
    0 & - \e_{\dot{\a}\dot{\b}} (\g^i)_{qp}  \cr } \right) ,
\qquad
(\tilde{\s}^{i})^{\underline{\a}\underline{\b}}
=
\left(\matrix{ - \e^{\a\b}(\g^i)_{qp} & 0 \cr
    0 & \e^{\dot{\a}\dot{\b}} (\tilde{\g}^i)^{qp}  \cr } \right) ,
\qquad
$$
where
$$
\s^a_{\a\dot{\b}} \equiv \e_{\a\b} \e_{\dot{\b}\dot{\a}}
(\tilde{\s}^a)^{\dot{\a}\b}  \qquad a=0,...,3
$$
are relativistic Pauli matrices,
$\e_{\a\b} = - \e_{\b\a}, ~~ \e_{12}= -1 = - \e^{12}$, and
$$ \g^i_{qp} = - \g^i_{pq} = - ((\tilde{\g}^i)^{qp})^{*} =
{1 \over 2} \varepsilon_{qprs} (\tilde{\g}^i)^{rs} $$
are Klebsh--Gordan coefficients for the group $SU(4)=SO(6)$ \cite{gsw}.

The worldvolume supervielbeins are
\begin{equation}\label{0.6}
 e^A = (e^a, e^{\hat{\alpha}})
 = dz^M e_M^A(z) =
 d\xi^{m} e_m^A + d\eta^{\hat{\mu}}  e_{\hat{\mu}}^A
\end{equation}
where
$
z^M = (\xi^{m}, \eta^{\hat{\mu}} )
~~$ ($m=0,...,p;~~ \hat{\mu}=1,...,16$) are local coordinates of the
worldvolume superspace of a super--$D$--$p$--brane,
$a$ is a $d= p+1$ tangent space vector index and $\alpha$ is a composite
16--valued spinor index of $SO(1, p) \times SO(9 - p)$.

Again in the $IIB$ case the representation with complex structure
\begin{equation}\label{0.7}
 e^{\hat{\alpha}} = ( e^{\a}_q , \bar{e}^{\dot{\a} q} )
\end{equation}
is convenient, where the indices $\a, \dot{\a}$ stand for spinor
representations of $SO(1,p)$  and $q$ is a spinor index of $SO(9-p)$.
For the Dirichlet $3$--brane
$$ p=3, \qquad \a=1,2, \qquad \dot{\a} = 1,2,
\qquad q =1,...,4. $$

To construct a super--$D$--brane action we introduce vector and
spinor Lorentz harmonics \cite{sok}--\cite{bpstv} given by
a $10\times 10$ matrix
$u_{\underline{a}}^{~\tilde{\underline{a}}}$ and
a $16\times 16$ matrix
$v_{\underline{\alpha}}^{~\tilde{\underline{\alpha}}}$
such that
\begin{equation}\label{1a}
 || u_{\underline{a}}^{~\tilde{\underline{ a}}} ||
 \qquad \in SO (1,9) \qquad
 \Rightarrow   \qquad
 u^{~\tilde{\underline{a}}}_{\underline{a}} \
\eta^{\underline{a}\underline{b}}
u^{~\tilde{\underline{b}}}_{\underline{b}}
= \eta^{\tilde{\underline{a}}\tilde{\underline{b}} }
\qquad and \qquad
||v_{\underline{\alpha}}^{~\tilde{\underline{\alpha}}}|| \in Spin (1,9)
\end{equation}
$u$ and $v$ matrices are related to each other by the following condition
of the invariance of the $\gamma$--matrices under the Lorentz rotations
 \begin{equation}\label{1b}
 u_{\underline{a}}^{~\tilde{\underline{a}}}
\sigma^{\underline{a}}_{\underline{\alpha}\underline{\beta}} = \
v_{\underline{\alpha}}^{\underline{\tilde{\alpha}}} ~
\sigma^{\tilde{a}}_{{\underline{\tilde{\alpha}}\tilde{\beta}}}
v^{~\underline{\tilde{\beta}}}_{\underline{\beta}}, \qquad
u^{~\underline{\tilde{a}}}_{\underline{a}}
\sigma^{\tilde{\underline{\alpha}}\tilde{\underline{\beta}}}
_{\underline{\tilde{a}}} =
v^{\underline{\tilde{\alpha}}}_{\underline{\alpha}} ~
\sigma^{\underline{\alpha\beta}}_{\underline{a}} ~
v^{\underline{\tilde{\beta}}}_{\underline{\beta}}.
\end{equation}

When adapted to the superbrane worldvolume,
$u_{\underline{a}}^{~\underline{\tilde{a}}}$ splits covariantly into
\begin{equation}\label{1c0}
u_{\underline{a}}^{~\underline{\tilde{a}}} =
(u_{\underline{a}}^a,
u^i_{\underline{a}}),
\end{equation}
where $u^a$ and $u^i$ are respectively
tangent and orthogonal vectors to the worldvolume
\footnote{The decomposition \p{1c0} and  \p{1c} are invariant under local
$SO(1,p) \times SO(9-p)$ transformations, which form a natural gauge
symmetry of the $p$--brane embedded into $D=10$ space--time. This provides
the possibility of treating Lorentz  harmonics \p{1c0} and \p{1c} as
coordinates of a coset space
${ {SO(1,9)} \over {SO(1,p) \times
SO(9-p) } }$ \cite{GIKOS,sok}--\cite{bpstv}.}.

In a similar way
$v^{~\tilde{
\alpha}}_{\underline{\alpha}}$ splits into
 \begin{equation}\label{1c}
 v^{~\underline{\tilde{\alpha}}}_{\underline{\alpha}} =
(v^{~\alpha }_{\underline{\alpha}q},
\bar{v}^{~\dot{\alpha}q}_{\underline{\alpha}} )
 \end{equation}
In the $p=3$ $IIB$ case $\a =1,2~,~ \dot{\a}=1,2,~q=1,...,4$
and bar denotes complex conjugation. The representation \p{0.51} can be
used to specify \p{1b} as follows $$ u^{~a}_{\underline{a}}
\s^{\underline{a}}_{\underline{\a}\underline{\b}} = v^{~
\a}_{\underline{\a}q} \s^a_{\a\dot{\b}}
\bar{v}^{~\dot{\a}q}_{\underline{\b} } +
v^{~ \a}_{\underline{\b}q} \s^a_{\a\dot{\b}}
\bar{v}^{~\dot{\a}q}_{\underline{\a} } ,
$$
\begin{equation}\label{1d}
u^{~i}_{\underline{a}} \s^{\underline{a}}_{\underline{\a}\underline{\b}}
=
v^{~\a}_{\underline{\a}q} \tilde{\g}^{i~qp}
v_{\underline{\b}\a p}-
\bar{v}^{~\dot{\a}q}_{\underline{\a}} \g^i_{qp}
\bar{v}^{~p}_{\underline{\b}\dot{\a}},
\end{equation}
$$
u^{~a}_{\underline{a}} \tilde{\s}_a^{\dot{\b}\a} \d^{~p}_{q} =
v^{~\a}_{\underline{\a}q} \tilde{\s}_{\underline{a}}
^{\underline{\a}\underline{\b}} \bar{v}^{~\dot{\b}p}_{\underline{\b}},
$$
$$
u^{~i}_{\underline{a}} \g^i_{qp} \e^{\a\b} =
v^{~\a}_{\underline{\a}q} \tilde{\s}_{\underline{a}}
^{\underline{\a}\underline{\b}}
v^{~\b}_{\underline{\b}p},
$$
$$
u^{~i}_{\underline{a}} \tilde{\g}^{i~qp} \e^{\dot{\a}\dot{\b}} =
- \bar{v}^{~\dot{\a}q}_{\underline{\a}} \tilde{\s}_{\underline{a}}
^{\underline{\a}\underline{\b}}
\bar{v}^{~\dot{\b}p}_{\underline{\b}}.
$$

The role of the
 Lorentz harmonics is to adapt the supervielbeins $E^{\underline{A}}$
 to the super--$p$--brane worldvolume \cite{bpstv} as follows
$$
E^{{\underline{A}}}       \qquad
\rightarrow \qquad
E^{\tilde{\underline{A}}} = ( E^{\tilde{\underline{a}}};
{\hat E}^{\hat{\tilde{\underline{\a}}}} )
$$
\begin{equation}\label{1.1}
E^{\tilde{\underline{a}}} \equiv
E^{\underline{a}}
~u^{~\underline{\tilde{a}}}_{\underline{a}} = (E^a, E^i)
\end{equation}

\begin{equation}\label{1.2}
{\hat E}^{\hat{\tilde{\underline{\a}}}} =
\cases{
(E^{\underline{\tilde{\a}}1}; E^2_{\underline{\tilde{\a}}}) = (
E^{\underline{{\a}}1}
v^{~\underline{\tilde{\a}}}_{\underline{{\a}}},
E^2_{\underline{{\a}}}
v_{\underline{\tilde{\a}}}^{\underline{{\a}}})
= (E^{\a 1}_q, E^{\dot{\a}q1}; E^{q2}_{\a}, E^{~2}_{\dot{\a}q})
& IIA \cr
(E^{\underline{\tilde{\a}}};
\bar{E}^{\underline{\tilde{\a}}}) =
(E^{\underline{{\a}}}
v^{~\underline{\tilde{\a}}}_{\underline{{\a}}},
\bar{E}^{\underline{{\a}}}
v^{~\underline{\tilde{\a}}}_{\underline{{\a}}})
=(E^{\a}_q, E^{\dot{\a}q};
\bar{E}^{\a}_q, \bar{E}^{\dot{\a}q})
& IIB \cr}
\end{equation}
where
\begin{equation}\label{1.2a}
E^{a} =
E^{\underline{{a}}}u^{~a}_{\underline{a}}, \qquad
E^{i} =
E^{\underline{{a}}}u^{~i}_{\underline{a}}, \qquad
\end{equation}
\begin{equation}\label{1.2b}
\matrix{ IIA : ~~&
E^{\a 1}_q =
E^{\underline{{\a}}1}
v^{~\a}_{\underline{\a}q},~~
E^{\dot{\a}q 1}=
E^{\underline{{\a}}1}
v^{~\dot{\a}q}_{\underline{\a}},~~
E^{q2}_{\a} =
E^2_{\underline{\a}}
v_{q\a}^{\underline{\a}}, ~~
E^{2}_{\dot{\a}q} =
E^2_{\underline{{\a}}}
v_{\dot{\a}q}^{\underline{\a}}, ~~
  \cr
IIB : ~~&
E^{\a}_q =
E^{\underline{{\a}}}
v^{~\a}_{\underline{\a}q} , ~~
E^{\dot{\a}q} =
E^{\underline{\a}}
\bar{v}^{\dot{\a}q}_{\underline{\a}} , ~~
\bar{E}^{\a}_q =
\bar{E}^{\underline{\a}}
v^{~\a}_{\underline{\a}q} , ~~
\bar{E}^{\dot{\a}q} =
\bar{E}^{\underline{\a}}
\bar{v}^{\dot{\a}q}_{\underline{\a}} \cr }.
\end{equation}

In addition to the superspace coordinates
$Z^{\underline{M}}$
super--$D$--$p$--branes
have a worldvolume (super) one-form
$$A=dz^{{M}} A_{{M}}(z^{{M}}).$$

In what follows the worldvolume supervielbeins \p{0.6} and \p{0.7}
will be regarded as ones induced by embedding and hence related to
pullbacks into the superworldvolume of \p{1.2a} and \p{1.2b} (see below).

\section{Original Super--$D$--$p$--brane actions}

In the $\kappa$--invariant formulation of refs.
\cite{c1}--\cite{bt} the worldvolume
${\cal M}_0$ is purely bosonic (not supersymmetric). Therefore
$\eta^{\hat{\mu}}$ and the
spinor vielbeins $e^{\hat{\alpha}}$ are absent and
 $$
 A = d\xi^m A_m (\xi)
$$ is a one--form
and $Z^M (\xi)$ are functions on ${\cal M}_0$.

The action functional obtained in \cite{c1}--\cite{bt} for
super--$D$--$p$--branes propagating in a background of $D=10$
type $II$ supergravity has the form
\begin{equation}\label{2}
S^{[1]} = I_{DBI} + I_{WZ}
\end{equation}
where $I_{DBI}$ is the Dirac--Born--Infeld action
\begin{equation}\label{3}
I_{DBI} = - \int_{{\cal M}_0} d^{p+1} \xi \sqrt{-det (g_{mn} +
e^{-{1\over 2} \phi} {\cal F}_{mn}} ),
\end{equation}
$\phi = \phi (Z^{\underline{M}} (\xi))$ is the dilaton superfield, $g_{mn}
= E^{\underline{a}}_m \eta_{\underline{a}\underline{b}}
E^{\underline{b}}_n$ is the induced metric and the field ${\cal F}_{mn}$
are the components of the 2--form
$$ {\cal F} = dA - B_{(2)}. $$
In (5) $B_{(2)}$  is the NS--NS (background) super 2--form
(see \cite{c0,c1,c2} and refs. therein) with the field
strength
\begin{equation}\label{4}
 H_{(3)} = dB_{(2)}
\end{equation}
The Wess--Zumino term $I_{WZ}$
\begin{equation}\label{5}
 I_{WZ} = - \int_{{\cal M}_0} {\cal L}^{WZ}_{p+1}
= \int_{{\cal M}_0} e^{\cal F} \wedge C
\end{equation}
is the integral over the worldvolume ${\cal M}_0$ of the Wess--Zumino
form ${\cal L}^{WZ}_{p+1}$ expressed in terms of the form ${\cal F}$ and
the formal sum of the RR super--n--forms $C$ ($n$ are even for the $IIB$
and odd for the $IIA$ case)
\begin{equation}\label{6} C = \oplus^9_{n=0}.
 C_{(n)}
 \end{equation}
 The field strengths of $C_{(n)}$ are
 \begin{equation}\label{7}
 R=
 e^{B_{(2)}} \wedge d(e^{-B_{(2)}} \wedge C) = \oplus^{10}_{n=1} R_{(n)}
 \end{equation}

 The invariance of the action \p{19} under fermionic
$\kappa$--transformations is stipulated by the existence of a $32\times
32$
traceless matrix $\bar\Gamma$ acting
on the tangent spinors and satisfying the condition $\bar\Gamma^2 =1$.
In refs.  \cite{c1,c2,bt} it was proved that this matrix exists for any
D--p--brane and is given by the formal sum
\begin{equation}\label{8}
d\xi^{p+1} \bar\Gamma = - {e^{{1\over 4} (p-3) \phi} \over L_{DBI}} exp
(e^{{-{1\over 2}\phi}} {\cal F}) \gamma \vert_{{\cal M}_0}
\end{equation}
 where
 $$ \cases{\gamma = \oplus_n \gamma^{(2n)} (K)^n I \quad {\rm \ in \ the \
IIB \ case} \cr \gamma = \oplus_n \gamma^{(2n+1)} \gamma^{11} \quad {\rm \
in \ the \ IIA \ case} \cr} $$
$$\gamma^{(n)} \equiv {1\over n!}
E^{\underline{a}_n}  ...  E^{\underline{a}_1}
\Gamma_{\underline{a}_1...{\underline{a}_n}}$$
and K and $I$ are the $2\times
2$ matrices given by Eq. \p{0.4} (or \p{0.5} in the real representation)
and $L_{DBI}\equiv \sqrt{-det(g + e^{{-\phi \over 2}}{\cal F})}$.

As usual $\Gamma_{\underline{a}}$ are the Dirac matrices in $D=10$
times
a charge conjugation matrix
(see \p{0.3} for the $IIB$ case)
and $\Gamma_{\underline{a}_1 ...{\underline
a_n}}$ is the antisymmetrized product of $\Gamma_{\underline{a}}$ with
unit weight.

The infinite reducible $\kappa$--symmetry transformations
of $Z^{\underline{M}}$ and $A(z)$ which leave the action \p{2} invariant
are
\begin{equation}\label{9} \delta_\k  Z^{\underline{M}}
E_{\underline{M}}^{\underline{a}} \equiv i_\k E^{\underline{a}}
= 0 , \qquad \delta_\k
Z^{{\underline{M}}}
\hat{E}_{\underline{M}}^{\hat{\underline {\alpha}}} \equiv
i_\k \hat{E}^{\hat{\underline {\alpha}}}
= \k^{\hat{\underline \alpha}}
\end{equation}
\begin{equation}\label{11}
\delta A = i_\k B_{(2)}
\qquad \Leftrightarrow \qquad  \delta {\cal F} = i_\k H_{(3)}
\end{equation}
where
\begin{equation}\label{91}
\k^{\hat{\underline \alpha}} =
\k^{\hat{\underline \beta}} (\bar\Gamma)^{~\hat{\underline
\alpha}}_{\hat{\underline \beta}}.
\end{equation}

For instance, for the 3--brane ${\cal L}^{WZ}_4$ and $\bar{\Gamma}$ are
given by
\begin{equation}\label{13}
{\cal L}^{WZ}_4 = \Bigl(C^{(4)}+{\cal
F} \wedge C^{(2)} + {1\over 2} {\cal F} \wedge {\cal F}
 C^{(0)})\vert_{{\cal M}_0},
 \end{equation}
 \begin{equation}\label{13.1}
 {\cal L}^{WZ}_5 \equiv d{\cal L}^{WZ}_4 = \Bigl(R^{(5)}+{\cal F} \wedge
R^{(3)} + {1\over 2} {\cal F} \wedge {\cal F} \wedge R^{(1)}),
 \end{equation}

 \begin{equation}\label{14}
 d^4\xi \bar{\Gamma} = - {1\over
 L_{DBI}} (\gamma^{(4)} + e^{-{\phi \over 2}}{\cal F} \wedge \gamma^{(2)}
K + {1\over 2} e^{-\phi }{\cal F} \wedge {\cal F} ) I \vert_{{\cal M}_0}.
\end{equation}

The superspace constraints for $H_{(3)}$, $R_{(5)}$, $R_{(3)}$ and
$R_{(1)}$ are
\begin{equation}\label{15a}
 H_{(3)} = e^{{\phi\over 2}} \Bigl[ {i\over 2}
 \hat{E}^{\hat{\underline \alpha}} \wedge
 \hat{E}^{\hat{\underline \beta}} \wedge E^{{\underline c}}
(\Gamma_{{\underline c}}
K)_{\hat{\underline{\alpha}}\hat{\underline{\beta}}} +
{1 \over 2} \hat{E}^{\hat{\underline \alpha}} \wedge E^{{\underline
b}} \wedge E^{{\underline c}} (\Gamma_{{\underline{c b}}}
K)_{\hat{\underline{\alpha}}}^{~\hat{\underline{\b}}}
\hat\Lambda_{\hat{\underline \beta}}\Bigr] + {1\over {3!}} E^{{\underline
a}} \wedge E^{{\underline b}} \wedge E^{{\underline c}} H_{{\underline {c
b a}}} \end{equation}
\begin{equation}\label{15b} R_{(5)} = {i \over 2}
 \hat{E}^{\hat{\underline \alpha}} \wedge \hat{E}^{\hat{\underline \beta}}
 \wedge E^{{\underline{c_3}}}
 \wedge E^{{\underline{c_2}}}
 \wedge E^{{\underline{c_1}}}
(\Gamma_{{\underline{c_1 c_2 c_3}}}  I)
_{\hat{\underline{\alpha}}
\hat{\underline{\beta}}}
+
{1\over 5!} E^{{\underline{a_5}}} \wedge ... \wedge E^{{\underline{a_1}}}
R_{\underline{a}_1...\underline{a}_5}
\end{equation}
\begin{equation}\label{15c}
 R_{(3)} = e^{-{\phi\over 2}} \Bigl[
 - {i \over 2} \hat{E}^{\hat{\underline \alpha}}
 \wedge \hat{E}^{\hat{\underline \beta}} \wedge E^{{\underline c}}
 (\Gamma_c J)_{\hat{\underline{\alpha}}
\hat{\underline{\beta}}} +
\hat{E}^{\hat{\underline \alpha}}
\wedge E^{{\underline b}} \wedge E^{{\underline c}}
(\Gamma_{{\underline{c b}}} K I)
_{\hat{\underline{\alpha}}}^{~\hat{\underline{\beta}}}
{\hat\Lambda}_{\hat{\underline \beta}}\Bigr]
+ {1\over 3!} E^{{\underline a}} \wedge E^{{\underline b}}
\wedge E^{{\underline c}}
R_{{\underline{c b a}}}
\end{equation}
\begin{equation}\label{15d}
R_{(1)} = 2 e^{-\phi } \hat{E}^{\hat{\underline \alpha}} (I{\hat\Lambda})
)_{\hat{\underline \alpha}} + E^{{\underline b}} R_{{\underline b}}
\end{equation}
where
\begin{equation}\label{16}
{\hat\Lambda}_{\hat{\underline\alpha}} = {1\over 2}
\nabla_{\hat{\underline \alpha}} \phi (Z) \end{equation} The $D=10$
$type~II$ supergravity torsion constraints are \begin{equation}\label{17}
 T^{{\underline a}} = {\cal D} E^{{\underline a}} =
 - {i \over 2} \hat{E}^{\hat{\underline \alpha}}
 \wedge \hat{E}^{\hat{\underline  \beta}}
 \Gamma^{{\underline a}}_{{\underline{\hat{\alpha}\hat{\beta}}}} =
- {i \over 2} (\hat{E} \Gamma^{{\underline a}} \hat{E})
 \end{equation}

\section{Super--$D$--$p$--brane action functional in
terms of differential forms} \vskip 0.3truecm

Instead of \p{2} we propose the following action functional
 \begin{equation}\label{18}
S = I_0 + I_{WZ} = \int_{{\cal M}_0} ({\cal L}^0_{p+1} + {\cal
L}^{WZ}_{p+1})
 \end{equation}
where the Wess--Zumino term $I_{WZ}$ is the same as before (Eq.\p{6}) and
\begin{equation}\label{19}
I_0 =  \int_{{\cal M}_0} {\cal L}^0_{p+1} \equiv
\int_{{\cal M}_0}  ({1 \over (p+1)!} E^{a_0} \wedge E^{a_1}
\wedge  ...  \wedge E^{a_p}
\epsilon_{a_0 a_1... a_p} e^{-{p-3 \over 2} \phi} \sqrt{-det (\eta_{ab} +
F_{ab})}
\end{equation}
$$
+ Q_{p-1} \wedge [e^{-{1\over 2} \phi} (dA - B_{(2)})
-{1\over 2} E^b \wedge E^a F_{ab}])
$$
Here $E^a$ are defined in \p{1.1}--\p{1.2b},
$F_{ab}$ is an auxiliary antisymmetric tensor field with tangent
space (Lorentz group) indices and $Q_{p-1}$ is a Lagrange multiplier which
produces the algebraic equation
\begin{equation}\label{20}
 F_{2} \equiv 1/2 E^b \wedge E^a F_{ab} = e^{-{\phi/over 2}}(dA - B_{(2)})
 \equiv e^{-{\phi/over 2}}{\cal F}
 \end{equation}
 and identifies the auxiliary field
 $F_{ab}$ with the components of the form ${\cal F}$ of the original
 action \p{3}.

The introduction into \p{18} of the term with $Q_{p-1}$ and the use of
$E^a$ \p{1.2a} enabled us to rewrite the DBI action \p{3} as the integral
of a differential $(p+1)$--form over ${\cal M}_0$ and, therefore, to
consider it on an equal footing with the WZ term. This explains why the
$\kappa$--variation of the DBI functional \p{3} is an integral of
a $(p+1)$--form \cite{c1,c2}.

The  Lagrange multiplier form $Q_{(p+1)}$  does not contain propagating
degrees of freedom because the variation of \p{19} with respect to the
auxiliary field $F_{ab}$ yields the equation
\begin{equation}\label{21}
Q_{p-1} \wedge E^b \wedge E^a = {1 \over (p+1)!} E^{a_0} \wedge E^{a_1}
 \wedge ...  \wedge E^{a_p} \epsilon_{a_0 a_1 ... a_p} e^{-{p-3 \over 2}
 \phi } {\partial \over \partial F_{ab}} (\sqrt{- det (\eta + F)})
\end{equation}
which is algebraic and can be easily solved
\begin{equation}\label{22}
Q_{p-1} = {1\over 4} \sqrt{-det (\eta + F)} e^{-{p-3 \over 2} \phi }
 E^{a_1}\wedge ... \wedge E^{a_{p-1}} \epsilon_{a_1 ... a_{p-1} ab} (\eta
+ F)^{-1 ab}.
\end{equation}

The variation of the Lorentz
harmonics $u^{~\underline{\tilde{a}}}_{\underline{a}}$
(contained in $E^a$) for getting field equations requires some
comments.  Since $u^{~\underline{\tilde{a}}}_{\underline{a}}$ must satisfy
the constraints \p{1a} one should add to the action \p{18} the term
$$ I_c = \int L_{\underline{\tilde{a}}\underline{\tilde{b}}}
(u^{\underline{\tilde{a}}}_{\underline{a}} \eta^{\underline{a b}}
u^{\underline{\tilde{b}}}_{\underline{b}}
- \eta^{\underline{\tilde{a}}\underline{\tilde{b}} }) $$
where $L_{\underline{\tilde{a}\tilde{b}}}$ are Lagrange multiplier
$(p+1)$-forms.
Then \p{18} would extend to  $S' = S+ I_c$. The field
equations $$ {\delta S' \over \delta u^i_{\underline{a}}}= 0 $$ lead to
$$ L^{ij} =0 = L^{ai}, $$
while the field equations
$$ u^{~a}_{\underline{a}} {\delta S' \over \delta u^{~b}_{\underline{a}}}
= 0 $$
specify $L_{ab}$ and
$$ u^i_{\underline{a}} {\delta S' \over \delta u^b_{\underline{a}}} = 0 $$
imply a so--called rheotropic condition \cite{bsv}
\begin{equation}\label{23}
 E^i \equiv dZ^{\underline{M}} E^i_{\underline{M}} = 0
\end{equation}
which reads that the pullback of $E^i$ into the worldvolume is zero.

Alternatively one can avoid adding the term $I_c$ but perform the variation
with respect to $u^{~\underline{\tilde{a}}}_{\underline{a}}$ according to
the rule
\begin{equation}\label{Om}
\delta u^{~\underline{\tilde{a}}}_{\underline{a}}=
u^{~\underline{\tilde{b}}}_{\underline{a}}~
\Omega^{~\underline{\tilde{a}}}_{\tilde{\underline{b}}} (\d )
\equiv
u^{~\underline{\tilde{b}}}_{\underline{a}}~
i_\d \Omega^{~\underline{\tilde{a}}}_{\tilde{\underline{b}}}
\end{equation}
where $
\Omega^{~\underline{\tilde{a}}}_{\tilde{\underline{b}}}
$ is the $SO(1,9)$--valued Cartan 1--form
\begin{equation}\label{Om1}
\Omega^{\tilde{\underline{a}}\tilde{\underline{b}}} = -
 \Omega^{\tilde{\underline{b}}\tilde{\underline{a}}} =
\left(\matrix{\Omega^{ab} & \Omega^{aj} \cr -\Omega^{bi} & \Omega^{ij} \cr}
\right)
\end{equation}
and $\Omega(\delta) = i_\delta \Omega$ (For the details see, for instance,
\cite{bzp,bpstv}). Then the variation of $S$ with respect to
$u^a_{\underline{a}}$ gives again \p{23}.
Taking into account Eq. \p{23} we get
\begin{equation}\label{24}
g_{mn} \equiv E^{\underline{a}}_m \eta_{\underline{ab}}
E^{\underline{b}}_n = E^a_m \eta_{ab} E^b_n
\end{equation}
so that $E^a_m$ can be regarded as induced worldvolume vielbeins.

Using the algebraic equation \p{20}, \p{22}, \p{23} and \p{24} we can
reduce the functional $I_0$ \p{19} to  $I_{DBI}$ \p{3}. This proves that
at the classical level the formulation under consideration is equivalent
to that of refs.  [1--4].

\section{$\kappa$--Invariance}

\bigskip

Since the action \p{18} is equivalent to \p{2}  its invariance under
$\kappa$--symmetry is guaranteed. However it is instructive, in view of
the consideration in the next section, to verify it explicitly.

Our action functional is the integral of a (p+1)--form ${\cal L}_{p+1}$
over the  world volume. If this form was the pullback of a target space
form its variation could be obtained
from the Lie derivative of the Lagrangian density ${\cal L}_{p+1}$
$$
\delta {\cal L}_{p+1} = i_\k d{\cal L}_{p+1} - d (i_\k {\cal L}_{p+1})
$$
so that, neglecting boundary terms, we would have
\begin{equation}\label{25} \delta S =
\int i_\k d {\cal L}_{p+1}.
\end{equation}

Note that Eq. \p{23} allows to identify the worldvolume vielbeins $e^a$
with a linear combination of the pullbacks of
$E^a$ tangent to the worldvolume.  The basic field variations defined by
contraction of the forms $E^a$ with the $\k$ parameter $ E^{a}(\d_\k )
\equiv i_\k E^a $ vanishes due to the definition of the $\kappa$--symmetry
\p{9}.  Hence the contraction of $Q$ vanishes as well (see \p{22}).

But since ${\cal L}_{p+1}$ also contains genuine worldvolume fields wich
are not the pullbacks of target space objects (such as Lorentz harmonics),
we must add to (47) the $\k$ variations of these fields.
However these
variations (which are still undefined) are
multiplied by the algebraic field equations \p{20}, \p{21}
and \p{23} and, therefore, they can be appropriately chosen to compensate
possible terms proportional to the algebraic equations that arise from the
variation of other terms.  It means, in particular, that when computing
$\delta S$ we can freely use these algebraic equations and, at the same
time, drop the $\k$ variations of these genuine worldvolume quantities if
we are not interested in their specific form.  Also notice that from
\p{11} the $\kappa$--variation of ${\cal F}$ is $$ \delta {\cal F} = i_\k
d {\cal F}= i_\k H $$ which can be also viewed as the Lie derivative of
${\cal F}$ provided we  formally assume $i_\kappa {\cal F} =0$.

 Thus in order to check the $\k$--invariance of $S$ one has  to compute
the differential of ${\cal L}_{p+1}^{WZ}$ and ${\cal L}^0_{p+1}$ (modulo
the algebraic equations \p{20}, \p{21} and \p{23}).

We shall explicitly compute the differential of ${\cal L}_{(p+1)}$ only
for the 3--brane.  The other cases can be treated in the same way.  From
\p{5} and the definition of the curvatures $R^{(n)}$ incoded in \p{7} and
the constraints \p{15a} -- \p{15d} one has
\begin{equation}\label{26}
d{\cal L}^{WZ}_4 = R_5 + {\cal F} \wedge R_3 + {1\over 2} {\cal F} \wedge
{\cal F}\wedge R_1 = {i \over 2} (\hat{E} \gamma^{(3)} \hat{E}) + (\hat{E}
\tilde\gamma^{(4)} \hat{\Lambda}),
\end{equation}
where
\begin{equation}\label{27}
\gamma^{(3)} = \Bigl[ {1\over {3!}} E^a \wedge
E^b \wedge E^c \Gamma_{cba} +  F_2 \wedge E^a \Gamma_a K \Bigr] I
\end{equation}
\begin{equation}\label{28}
\tilde \gamma^{(4)} = \Bigl[ F_2
\wedge E^a \wedge E^b \Gamma_{ba} K - F_2 \wedge F_2 \Bigr] I
\end{equation}
and
$$ \G^a \equiv \G^{\underline{a}} u^a_{\underline{a}}.
$$

On the other hand the differential of ${\cal L}_0$ is
$$
d {\cal L}^0_4 =
\sqrt{-det (\eta + F)} \{ {1\over {3!}} \epsilon_{a_1 a_2 a_3 a} E^{a_1}
\wedge E^{a_2} \wedge E^{a_1} \wedge T^{a} -
$$
$$
- {1\over 4} E^{a_1}
\wedge E^{ a_2} \epsilon_{a_1 a_2 a_3 a_4} ((\eta + F)^{-1})^{a_3 a_4}
 \wedge [e^{-{\phi \over 2}} H_3 + {1\over 2} F_2 \wedge d\phi + T^a
 \wedge E^b F_{ba}] \},
$$ where $T^a \equiv
 T^{\underline{a}}u^a_{\underline{a}}$.  Using the constraints \p{15a} and
\p{17} and making some algebraic manipulations we can rewrite $d{\cal
L}^0_4$ as
$$
d {\cal L}^0_4  = \sqrt{- det (\eta + F)} [ - {i\over
{2~3!}} \epsilon_{a_1 a_2 a_3 b} E^{a_1} \wedge E^{a_2} \wedge E^{a_3}
\wedge  \Bigl(\hat{E}\G_a ((\eta + K F)^{-1})^{ab} \hat{E}\Bigr) +
$$
$$ +
\epsilon_{a_1 a_2 a_3 a_4} E^{a_1} \wedge ... \wedge E^{a_4} \wedge \{ {1
\over 16 } \Bigl(\hat{E}\G_b ((\eta + K F)^{-1})^{ba}
\G_a\hat{\Lambda}\Bigr) - {1 \over 4} (\hat{E}\hat{\Lambda}) \} ],
$$
where
$$ (\eta + K F)^{-1ba} \equiv (\eta +  F)^{-1\{ ba \} } 1  + (\eta +
 F)^{-1ba} K = $$ $$ (\eta^{ba} - F^b_{~c}F^{ca}+...) 1 + (F^{ba} -
 F^{bc}F_{cd}F^{da}+...) K
 $$
 Inserting the unit matrix $\bar\Gamma^2$ and
using the remarkable identity
$$ \bar\Gamma \Gamma_a = {1 \over \sqrt{-
det (\eta + F)}} [{1\over {3!}} \Gamma_{a_1 a_2 a_3} + {1\over 2} F_{a_1
a_2} \Gamma_{a_3} K] \epsilon^{a_1 a_2 a_3 b} (\eta + FK)_{ba}
$$
one gets
\begin{equation}\label{29}
d {\cal L}^0_4 =-{i\over2}(\hat{E} \bar\Gamma
 \gamma^{(3)} \hat{E})  - (\hat{E} \bar \Gamma \tilde\gamma^{(4)}
 \hat{\Lambda})
 \end{equation}
 $\gamma^{(3)}$ and $\tilde\gamma^{(4)}$ are
 given in \p{27} and \p{28}.

In conclusion
\begin{equation}\label{30}
d{\cal L}_4 = d{\cal L}^0_4 + d
 {\cal L}^{WZ}_4 =  i(\hat{E}^{(-)} \gamma^{(3)} \hat{E}^{(-)}) -2
(\hat{E}^{(-)} \tilde\gamma^{(4)} \hat{\Lambda}),
\end{equation}
where
\begin{equation}\label{31} \hat{E}^{(-)\hat{\underline\alpha}} = {1\over
2}(\hat{E}(1 - \bar\Gamma))^{\hat{\underline\alpha}}.
\end{equation}
(Going from \p{26} and \p{28}  to \p{30} we used the properties
$\bar\Gamma^T=-\bar\Gamma$ and
$\bar\Gamma\gamma^{(3)}=-\gamma^{(3)}\bar\Gamma^T$).

At this point the $\kappa$--invariance of $S$ becomes obvious (see \p{9},
\p{91} and \p{31}) since
\begin{equation}\label{32} i_\k
 \hat{E}^{(-)\hat{\underline\alpha}} =0 = i_k E^{{\underline a}}
 \end{equation}
 and hence
 \begin{equation}\label{33}
 i_k d{\cal L} =0 .
\end{equation}

\section{Irreducibility of $\kappa$ symmetry in the Lorentz harmonic
formulation}

It should be stressed that the use of the Lorentz harmonics
provides us with the possibility of  extracting
the covariant set of $16$ independent parameters of the
$\kappa$--symmetry.  This means that passing from the original  functional
\p{2}, \p{3}, \p{5} to the classically equivalent  action \p{18}, \p{19},
\p{5} we achive an {\sl irreducible description} of the $\k$--symmetry
(see \cite{bzp,stv}--\cite{dpol} for superparticles, superstrings and
type $I$ super--p--branes).

As an example let us consider the $3$--brane case.
In 32--component spinor notations used in the previous Section
the ${ SO(1,9) \over SO(1,3) \times SO(6)}$ Lorentz harmonics \p{1b}
are represented by the reducible matrix
\begin{equation}\label{5.1}
v_{\hat{\underline{\a}}}^{~\hat{\tilde{\underline{\a}}}} =
\left(\matrix{
v_{\underline{\a}}^{~{\underline{\tilde{\a}}}} & 0 \cr
0 & v_{\underline{\a}}^{~{\underline{\tilde{\a}}}}  \cr } \right).
\end{equation}

To extract the irreducible part from the $\kappa$--symmetry parameter
\p{91}
\begin{equation}\label{5.2}
\k^{\hat{\underline{\a}}} = i_{\d }
\hat{E}^{(+)\hat{\underline{\a}}} =
i_{\d } \hat{E}^{\hat{\underline{\b}}} ({ 1
+\bar{\Gamma} \over 2})_{\hat{\underline{\b}}}^{~\hat{\underline{\a}}}
\end{equation}
it is necessary first of all
to contract $\k^{\hat{\underline{\a}}}$ with the $32\times 32$
Lorentz--harmonic matrix \p{5.1}. As a result we get the parameter
\begin{equation}\label{5.3}
\k^{\hat{\tilde{\underline{\a}}}}
= \k^{{\hat{\underline{\b}}}}
v_{{\hat{\underline{\b}}}}
^{~\hat{\tilde{\underline{\a}}}} ~=~
(\k^{\tilde{\underline{\a}}};
\bar{\k}^{\tilde{\underline{\a}}}) ~= ~~
(\k^\a_q, \k^{\dot{\a}q} ;
\bar{\k}^\a_q, \bar{\k}^{\dot{\a}q})
\end{equation}
covariantly splitted into four pieces.

The set of the parameters
$\k^{\hat{\tilde{\underline{\a}}}}$ \p{5.3} satisfies the condition
\begin{equation}\label{5.4}
\k^{\hat{\tilde{\underline{\a}}}}
= \k^{\hat{\tilde{\underline{\b}}}}
(\bar{\Gamma}^{\prime})_{\hat{\tilde{\underline{\b}}}}
^{~\hat{\tilde{\underline{\a}}}}
\end{equation}
where (in the complex representation \p{0.2})
$$
(\bar{\Gamma}^{\prime})
_{\hat{\tilde{\underline{\b}}}}
^{~\hat{\tilde{\underline{\a}}}}
\equiv
v_{\hat{\tilde{\underline{\b}}}}
^{~\hat{{\underline{\b}}}}
\bar{\Gamma}_{{\hat{\underline{\b}}}} ^{~{\hat{\underline{\a}}}}
v_{{\hat{\underline{\a}}}}
^{~\hat{\tilde{\underline{\a}}}} =
$$
\begin{equation}\label{5.7}
{ 1 \over {\sqrt{-det(\eta + F)}}}
\left(\matrix{
(\s_{(4)})
_{\tilde{{\underline{\b}}}}^{~\tilde{{\underline{\a}}}}
 + {i \over 8} \epsilon^{abcd}F_{ab}F_{cd}
 \d
 _{\tilde{{\underline{\b}}}}^{~\tilde{{\underline{\a}}}}
 & -2i (\s_{(2)})
 _{\tilde{{\underline{\b}}}}^{~\tilde{{\underline{\a}}}}
 \cr
   2i (\s_{(2)})
 _{\tilde{{\underline{\b}}}}^{~\tilde{{\underline{\a}}}}
    &
- (\s_{(4)})
 _{\tilde{{\underline{\b}}}}^{~\tilde{{\underline{\a}}}}
 - {i \over 8} \epsilon^{abcd}F_{ab}F_{cd} ~
 \d_{\tilde{{\underline{\b}}}}^{~\tilde{{\underline{\a}}}}
  \cr} \right)
\end{equation}
$$
(\s_{(4)})_{\tilde{{\underline{\a}}}}^{~\tilde{{\underline{\b}}}}
= {i\over 4} \e^{abcd} (\s_{abcd})
_{\tilde{{\underline{\a}}}}^{~\tilde{{\underline{\b}}}} =
$$
$$
\left(\matrix{
              \d^{~\b}_{\a} \d^q_{~p} & 0 \cr
              0 & -\d^{~\dot{\b}}_{\dot{\a}} \d_q^{~p}  \cr }  \right)
$$
$$
(\s_{(2)})_{\tilde{{\underline{\a}}}}^{~\tilde{{\underline{\b}}}}
= {1\over 8} \e^{abcd} F_{ab} (\s_{cd})
_{\tilde{{\underline{\a}}}}^{~\tilde{{\underline{\b}}}} =
\left(\matrix{
{i\over 4} F^{ab} (\s_a\tilde{\s}_b)^{~\b}_{\a} \d^q_{~p} & 0 \cr
 0 & -{i\over 4} F^{ab} (\tilde{\s}_a \s_b )^{\dot{\b}}_{~\dot{\a}}
      \d_q^{~p}  \cr} \right).
$$

 The solution of Eq. \p{5.4} has
the form
\begin{equation}\label{5.8}
\k^{\hat{\tilde{\underline{\a}}}} = i_{\d}
E^{(+)\hat{\tilde{\underline{\a}}}}
= \left( \matrix{
\k^\a_q   \cr
 2 b_{-} \bar{\k}^{\dot{\b}q} f^{~\dot{\a}}_{\dot{\b}}  \cr
2 b_{+} \k^\b_q f^{~\a}_{\b} \cr
\bar{\k}^{\dot{\a}q}\cr } \right) ^{T}
\end{equation}
where
\begin{equation}\label{5.9}
f_{\a\b} = f_{\b\a} =
{1\over 4}   F^{ab} (\s_a \tilde{\s}_b)_{\a\b} ,
\qquad
\bar{f}_{\dot{\a}\dot{\b}} =
\bar{f}_{\dot{\b}\dot{\a}} =
{1\over 4}   F^{ab} (\tilde{\s}_a
\s_b)^{\dot{\b}}_{~\dot{\a}} ,
\end{equation}
$$
b_{\pm} = { 1 \over {1 \pm {i\over 8} \e^{abcd} F_{ab} \s_{cd}
+ \sqrt{-det(\eta + F)}}} .
$$

The solution \p{5.8} singles out 16 independent covariant  parameters
$\k^{\a}_q,~\bar{\k}^{\dot{\a}q}$ of the {\sl irreducible} $\k$--symmetry
of the super--$D$--$3$--brane.

\bigskip

Remember that the infinite reducibility of the $\kappa$--symmetry
in the Green--Schwarz formulation of superstring theory
\cite{gsw} is a main problem which hampers the covariant quantization.
The same problems
appear in the DBI like formulation of the super--$D$--$p$--branes
\cite{c1}--\cite{bt}, which have $\k$--symmetry realized in the
infinitely reducible form.

In this respect it is remarkable that, as has been proved in this Section,
the $\kappa$--symmetry of the Lorentz--harmonic formulation of the
super--$D$--$p$--branes is realized in irreducible form.

\section{Generalized action functional and superfield equations of motion}

The action \p{18}, \p{19} and \p{5} is written in
terms of differential forms without any use of Hodge operation $~*~$
and, hence, can be used for the construction of the generalized action
\cite{bsv} (see \cite{rheo} for supergravity) for Dirichlet
super--p--branes in a $D=10$ $type~II$ supergravity background.

This generalized action can be regarded as a dynamical basis for deriving
superfield equations of motion of the
super--$D$--$p$--branes as geometrical conditions of embedding their
superworldvolumes into a target superspace  \cite{bpstv,bsv,hs1,hs2}.

Suppose the integration surface in the functional \p{18} to be
an arbitrary surface
\begin{equation}\label{6.1}
{\cal M}^{p+1} = \{ ( \xi^{m}, \eta^{\hat{\mu}}(\xi) )\}
\end{equation}
in a worldvolume superspace
\begin{equation}\label{6.2}
\Sigma^{(p+1 \vert 8+8)} = \{ ( \xi^{m}, \eta^{\hat{\mu}} )\}
\end{equation}
of the $type~II$ super--$p$--brane specified by 16 Grassmann functions
(Goldstone fermions \cite{va})$$
\eta^{\hat{\mu}} = \eta^{\hat{\mu}}(\xi^m) .
$$
Henceforth suppose all the coordinates of the target superspace and
Lorentz harmonics involved into \p{19} and \p{5} to be superfields on
$\Sigma^{(p+1 \vert 8+8)}$
\begin{equation}\label{6.3}
Z^{\underline{M}} = Z^{\underline{M}} (\xi , \eta ) , \qquad
u^{~\underline{a}}_{\underline{m}} =
u^{~\underline{a}}_{\underline{m}} (\xi , \eta ) , ~~~ ...
\end{equation}
but restricted to an arbitrary surface ${\cal M}^{(p+1)}: ~~ \eta = \eta
(\xi ) $
\begin{equation}\label{6.4} Z^{\underline{M}} = Z^{\underline{M}}
(\xi , \eta (\xi) ) , \qquad u^{~\underline{a}}_{\underline{m}} =
u^{~\underline{a}}_{\underline{m}} (\xi , \eta (\xi ) ) , ~~~ ...
\end{equation}
In this way we get a generalized action for super-D-p-branes
(see \cite{bsv} for superstrings and type $I$ super--p--branes)  in $D=10$
type $II$ supergravity background
 \begin{equation}\label{6.5}
S = \int_{{\cal M}^{p+1}=\{ (\xi, \eta = \eta (\xi ) ) \} }
({\cal L}^0_{p+1} + {\cal L}^{WZ}_{p+1}), \end{equation}
$$
{\cal L}^0_{p+1} \equiv
 ({1 \over (p+1)!} E^{a_0} \wedge E^{a_1} \wedge  ...
\wedge E^{a_p} \epsilon_{a_0 a_1... a_p} e^{-{p-3 \over 2} \phi}
\sqrt{-det (\eta_{ab} + F_{ab})} $$
$$ + Q_{p-1} \wedge
[e^{-{1\over 2} \phi} (dA - B_{(2)}) -{1\over 2} E^b \wedge E^a F_{ab}])
\vert_{{\cal M}^{p+1}},
$$
$$
 {\cal L}^{WZ}_{p+1}
= e^{\cal F} \wedge C \vert_{{\cal M}^{p+1}}.
$$
In \p{6.5} the formal sum of the forms $C$ is defined by \p{6} and
all the variables should be regarded as superfields
\p{6.3} restricted to the bosonic surface ${\cal M}^{p+1}$ \p{6.4}.
All the forms are defined on the whole worldvolume superspace
\p{6.2}
and pulled back into ${\cal M}^{p+1}$ (this is denoted by
$\vert_{{\cal M}^{p+1}}$).
 For example, the external differential is
\begin{equation}\label{6.6}
d = d\xi^m \partial_m + d\eta^{\hat\mu} \partial_{\hat\mu} =
e^A \nabla_A = e^a \nabla_a + e^{\hat{\a}} \nabla_{\hat{\a}} ,
\end{equation}
and its pullback is
\begin{equation}\label{6.7}
d = d\xi^m (\partial_m + \partial_m \eta^\mu (\xi ) \partial_\mu )
 = d\xi^m (e^A_m  + \partial_m \eta^\mu (\xi ) e_\mu^{A}) \nabla_{A} .
 \end{equation}

Note that the difference in construction of the generalized action
\p{6.5} from the generalized action for ordinary type I super--p--branes
\cite{bsv} is that \p{6.5}  does not contain intrinsic worldvolume
supervielbeins \p{0.6} and \p{0.5} as independent auxiliary fields
\footnote{In this sense the action \p{6.5} is closer to a formulation of
bosonic strings with auxiliary vector fields proposed in \cite{vzUFZh}
than to the Lorentz--harmonic formulation of refs. \cite{bzp,bpstv,bsv}.}.
As we shall see below and in the next Section worldvolume supergeometry is
induced and completely specified by conditions of embedding into target
superspace.

A reason why one finds more convenient to construct the action without
intrinsic worldvolume supervielbeins is
the presence of the worldvolume 1--form gauge (super)field and the
nonlinear nature of the $D$--brane theories reflected in the form
of the DBI functional in the original formulation [1-5].

The generalized action principle \cite{rheo,bsv} is based on the
 requirement that the equations of motion originate from the
 vanishing of the variation of the functional \p{6.5} with respect
 to the variation of the superfields involved
 as well as under arbitrary variations of the surface ${\cal M}^{(p+1)}$
 itself which can be regarded as a variation with respect to the Goldstone
 fermion field $\eta^{\hat{\mu}}(\xi )$.

 For the Lagrangian form under consideration it can be proved (see
Refs. \cite{bsv} ) that the variation with respect to the
surface ${\cal M}^{(p+1)}$, i.e.
$${\d S \over {\d \eta (\xi)}} = 0 ,$$
does not lead to new equations of motion. (The letter are consequences of
the equations of motion for other fields).  This implies the
superdiffeomorphism invariance of the generalized action \cite{rheo,bsv}.

In the same way as it was done in refs. \cite{bsv} for superstrings
and $type~I$ super--$p$--branes one can show that this superdiffeomorphism
invariance is related to the irreducible $\kappa$--symmetry of the
Lorentz--harmonic formulation \p{18} as well as to the infinitely reducible
$\kappa$--symmetry \p{9} -- \p{91} of the original DBI--like
formulation.
Thus, as in the case of ordinary super--p--branes (see
\cite{bsv}--\cite{bers94}), the generalized action clarifies the origin of
the $\kappa$--symmetry of the super--$D$--$p$--brane theories as a
manifestation of the local worldvolume supersymmetry.

The fact that the surface ${\cal M}^{(p+1)}$ \p{6.1} is arbitrary
and that the whole set of such surfaces spans the whole worldvolume
superspace \p{6.2} ensures the possibility of considering
equations of motion ($\d S / \d Z^{\underline{M}} = 0$, etc. )
as  superfield ones, i.e.  as equations for the superforms and superfields
defined in the whole worldvolume superspace $\Sigma^{(p+1\vert 8+8)}$.
These equations are formally the same as ones obtained from the
action \p{18} (see, for instance, \p{23}, \p{20} and \p{21}):
\begin{equation}\label{6.8}
 E^i \equiv dZ^{\underline{M}} E^i_{\underline{M}} = 0 ,
\end{equation}

\begin{equation}\label{6.9}
 i \gamma^{(p)}_{\hat{\underline{\a}} \hat{\underline{\b}}}
\hat{E}^{(-)\hat{\underline{\b}} }
- ({1\over 2}(1-\bar{\Gamma}) \tilde\gamma^{(p+1)})_{\hat{\underline{\a}} }
^{~~\hat{\underline{\b}} }~
\hat{\Lambda}_{\hat{\underline{\b}}} = 0 ,
\end{equation}
but now these are the equations for superforms and hence they should be
expanded in the whole basis \p{0.6} of the supervielbeins
\begin{equation}\label{6.10}
 e^A = (e^a, e^{\hat{\alpha}})= (e^a, e^{\alpha}_q, {\bar e}^{\dot{\a}q})
\end{equation}
of the worldvolume superspace \p{6.2}, the external differential
being determined in \p{6.6}.

As a result \p{6.8} contains now a spinor component
\begin{equation}\label{6.11}
E^i \equiv dZ^{\underline{M}} E^i_{\underline{M}}  =
e^a E_a^{~i} + e^{\hat{\a}} E^i_{\hat{\a}} = 0  \qquad
\Rightarrow \qquad E^{~i}_{{a}} = 0, \qquad E^{~i}_{\hat{\a}} = 0.
\end{equation}
The vanishing of the vector component of \p{6.8} ($E^i_a = 0$) implies that
the worldvolume bosonic vielbein superform can be identified with the
induced vielbein $E^a$ up to a
nonsingular matrix $m_b^{~a} \equiv E_b^{~a}$ \footnote{The choice of the
matrix $m_b^{~a}$ is a matter of convenience and can be used to get the
main spinor--spinor component of the worldvolume torsion in the standard
form (see below).}
\begin{equation}\label{6.12} e^b m_b^{~a} = E^a  \qquad \Rightarrow
\qquad E^{~a}_{\hat{\a}} = 0 ,
\end{equation}
(which is a conventional
rheotropic condition \cite{bsv}). Eqs. \p{6.11} and \p{6.12} result in the
geometrodynamic condition \begin{equation}\label{6.13}
E^{~\underline{a}}_{\hat{\a}} =
0
\end{equation}
being a basic point of the superfield (twistor--like)
description of superstrings and the $type~I$ superbranes  \cite{stv} --
\cite{bers94}, \cite{bpstv} as well as of the $D=11$
super--$5$--brane \cite{hs2} and Dirichlet--p--branes in the linearized
approximation   \cite{hs1}.

Decomposing Eq. \p{6.9} in the basic worldvolume $(p+1)$--forms we find
that the second term contains the input proportional to the form
$e^{a_1}\wedge ... \wedge e^{a_{p+1}} \e_{a_1...a_{p+1}}$
only, while the
input proportional to the basic form
$e^{a_1}\wedge ... \wedge e^{a_{p}} \wedge e^{\hat{\a}} $ comes from the
first term which gives rise to
an independent geometrical equation for the components of the Grassmann
supervielbein $E^{\hat{\underline{\a}}}$
(a fermionic {\sl rheotropic condition} \cite{bsv}). Upon omitting a
nonsingular matrix multiplier one reduces this equation to
\begin{equation}\label{6.14}
\hat{E}^{(-)\hat{\underline{\a}}}_{\hat{\a}} = 0,
\end{equation}
or, in terms of differential forms, to
\begin{equation}\label{6.15}
\hat{E}^{(-)\hat{\underline{\a}}} = e^a
\hat{\psi}_a^{(-)\hat{\underline{\a}}}.
\end{equation}

{\bf Eq.\p{6.14} (or \p{6.15}) together with \p{6.8} and \p{6.12} form the
complete set of superfield equations for the super--D--p--branes
in $D=10$ $type~II$ supergravity background.}

The superfield equation being the coefficient of
the basic $(p+1)$--form
$e^{a_1}\wedge ... \wedge e^{a_{p+1}} \e_{a_1...a_{p+1}}$ in \p{6.9}
expresses the gamma trace of the  superfield
$\hat{\psi}_a^{(-)\hat{\underline{\a}}}$ \p{6.15} through the derivatives
of the dilaton superfield $\Lambda_{\hat{\underline{\a}}}  = 1/2 {\cal
D}_{\hat{\underline{\a}}}  \phi $.
For example, for the case of $3$--brane we get
$$
\s^a_{\a \dot{\a}} \bar{\psi}_a^{(-)\dot{\a}q} =
4i (\sqrt{-det(\eta + F)}-1) \bar{\Lambda}_\a^q
+ 8i { {b_+} \over  {b_- }} f_\a^{~\b}
\Lambda^q_\b ,
$$
$$
\psi^{(-){\a}}_{a~q} \s^a_{\a \dot{\a}} =
- 4i (\sqrt{-det(\eta + F)}-1) \Lambda_{\dot{\a}q}
- 8i { {b_+} \over  {b_- }} {\bar f}_{\dot\a}^{~\dot\b}
\bar{\Lambda}_{\dot{\b}q} .
$$
 This equation is the same as the
equation of motion of $\Theta^{\hat{\underline\mu}}$ derived from the
``component'' action \p{18}, but with worldvolume superfields instead of
fields.  We should stress that these fermionic equations can be obtained
from the selfconsistency conditions for Eqs. \p{6.8}, \p{6.12} and
\p{6.15}.

\section{Superfield equations for $type~IIB$ super--D--3--brane}

As an instructive example let us
consider the superfield equations for a
{\bf super--D--3--brane} in {\bf flat} $D=10$ $N=IIB$ superspace.

As it was done for the parameter of the
$\kappa$--symmetry in the section {\bf 4},
using the explicit form of the projector $\bar{\Gamma}$ \p{5.7} in the
complex representation \p{0.2} we can express two independent
8--component forms $E^{(-)\dot{\a}q}$ and $\bar{E}^{(-)\a}_{~~~q} $
of the 32--component form
\begin{equation}\label{6.150}
\hat{E}^{(-)\tilde{\hat{\underline{\a}}}} =
\hat{E}^{(-){\hat{\underline{\b}}}} v_{{\hat{\underline{\b}}}}
^{~\tilde{\hat{\underline{\a}}}} =
 \left( \matrix{
 2 b_{+} \bar{E}^{(-)\b}_{~~~q} f^{~\a}_{\b} \cr
 E^{(-)\dot{\a}q} \cr
 \bar{E}^{(-)\a}_{~~~q} \cr
 2 b_{(-)} E^{(-)\dot{\b} q}
\bar{f}^{~\dot{\a}}_{\dot{\b}}  \cr } \right) ^{T}
\end{equation}
(for the definition of $f,~\bar f$ and $b_{\pm}$ see \p{5.9})
in terms of the covariant components
$
E^{\a}_{q},
 E^{\dot{\a}q},
 \bar{E}^{\a}_{q},
$
and
$
 \bar{E}^{\dot{\a} q}
$
of the complete pullback into the worldvolume superspace of the Grassmann
vielbein 1--form
$$
\hat{E}^{\hat{\tilde{\underline{\a}}}} =
\hat{E}^{{\hat{\underline{\b}}}} v_{{\hat{\underline{\b}}}}
^{~\hat{\tilde{\underline{\a}}}}
= \left( \matrix{
E^{\a}_{q} \cr
 E^{\dot{\a}q} \cr
 \bar{E}^{\a}_{q} \cr
 \bar{E}^{\dot{\a} q}
\cr } \right) ^{T}.
$$

In this way we get
\begin{equation}\label{6.16}
E^{(-)\dot{\a}q}  = { 1 \over {2b_{-} \sqrt{-det(\eta+F)}}}
(E^{\dot{\a}q}    - 2 b_- \bar{E}^{\dot{\b}q}
\bar{f}^{~\dot{\a}}_{\dot{\b}} ),
\end{equation}
$$
\bar{E}^{(-)\a}_{~~~q}
= { 1 \over {2b_{+} \sqrt{-det(\eta+F)}}} (\bar{E}^{\a}_{q} - 2 b_+
E^{\b}_{q} f^{~\a}_{\b} )
$$
In flat $D=10$ $IIB$ superspace (where, in particular,
$\Lambda_{\hat{\underline{\a}}} = 0$) after some algebra Eq.
\p{6.9} for the super--D--$3$--brane takes the form
\begin{equation}\label{6.17}
\hat{E}^{(-)}\wedge \gamma^{(3)} = 0 \qquad
\Leftrightarrow \qquad
\cases{
{1 \over 3!} E^{a_1}\wedge E^{a_2} \wedge  E^{a_3} \wedge
\e_{a_1 a_2 a_3 a_4} ~m^{~a_4}_{a} \wedge \bar{E}^{(-)\a}_{~~~q}
\s^a_{\a\dot{\a}} = 0 , \cr
{1 \over 3!} E^{a_1}\wedge E^{a_2} \wedge  E^{a_3} \wedge
\e_{a_1 a_2 a_3 a_4} ~m^{~a_4}_{a} \wedge
\s^a_{\a\dot{\a}} E^{(-)\dot{\a}q}= 0 }
\end{equation}
In \p{6.17}
\begin{equation}\label{6.18}
m_a^{~b} = \d_a^{~b} + b_+ b_- Sp(\tilde{\s}_a f \s^b \bar{f})
\end{equation}
where $f,~\bar{f}$ are spinor representations for the self--dual and the
anti--self dual part of the tensor $F_{ab}$ defined in Eq.\p{5.9} together
with $b_{\pm}$.  Choosing the worldvolume vielbein as in \p{6.12} with the
matrix $m$ given by \p{6.18} we get from \p{6.17}
\begin{equation}\label{6.19}
\cases{ {1 \over 3!} e^{a_1}\wedge e^{a_2}
\wedge  e^{a_3} \wedge \e_{a_1 a_2 a_3 a} \wedge \bar{E}^{(-)\a}_{~~~q}
\s^a_{a\dot{\a}} = 0 , \cr
{1 \over 3!} e^{a_1}\wedge e^{a_2} \wedge  e^{a_3} \wedge
\e_{a_1 a_2 a_3 a_4} \wedge
\s^a_{\a\dot{\a}} E^{(-)\dot{\a}q}= 0 }
 \end{equation}

Using the expressions \p{6.16} we can
represent the geometrical Grassmann equations ({\sl rheo\-tro\-pic
conditions}) \p{6.15} in the form
$$ \bar{E}^{\a}_{q}  = 2 b_+ E^{\b q}
f^{~\a}_{\b} + e^a \psi_{aq}^{~\a} ,
$$
\begin{equation}\label{6.20}
E^{\dot{\a}q}  = 2 b_- \bar{E}^{\dot{\b}q} \bar{f}^{~\dot{\a}}_{\dot{\b}}
+ e^a
\bar{\psi}_a^{\dot{\a}q} ,
\end{equation}
where (see \p{6.15})
$$
\psi_{aq}^{~\a}
= 2 b_- \sqrt{det(\eta +F)} \bar{\psi}_{a~q}^{(-)\a} ,
\qquad
\bar{\psi}_a^{\dot{\a}q}
= 2 b_+ \sqrt{det(\eta +F)} \psi_{a}^{(-)\dot{\a}q} .
$$
Equations \p{6.20} together with \p{6.11} and \p{6.12}
\begin{equation}\label{6.11'}
E^i \equiv dZ^{\underline{M}} E^{\underline{a}}_{\underline{M}}
u_{\underline{a}}^i  = 0 ,
\end{equation}
\begin{equation}\label{6.12'}
E^a \equiv dZ^{\underline{M}} E_{\underline{M}}^{~\underline{a}}
u_{\underline{a}}^{~a}
= e^b m_b^{~a}   \qquad
\end{equation}
form the complete set of the superfield equations ({\sl rheotropic
conditions} \cite{bsv}) for the $type~IIB$ super--3--brane in  flat
$D=10$, $N=2B$ superspace.

To completely specify the worldvolume superspace geometry we should
add to the above equations {\sl conventional rheotropic conditions}
determining the Grassmann worldvolume supervielbeins
$e^{\hat{\a}}= (e^{\a}_q, \bar{e}^{\dot{\a}q})$ which are not present in
the generalized action.  It can be proved (in a way similar to one
described in the refs. \cite{bsv} for superstrings and ordinary
super--$p$--branes) that $e^{\hat\alpha}$  can be identified with a linear
combination of  $E^{\a}_q$, $\bar{E}^{\dot{\a}q}$ and $E^a$
pulled back into superworldvolume
\begin{equation}\label{6.20'}
E^{\a}_q - E^a \chi_{aq}^{~\a}= e^{\a}_q ,
\qquad
\bar{E}^{\dot{\a}q} - E^a \bar{\chi}_{a}^{\dot{\a}q}
= \bar{e}^{\dot{\a}q} ,
\qquad
\end{equation}

The dynamical equations of motion of $\Theta^{\underline\mu}$ contained in
\p{6.19}
\begin{equation}\label{6.21}
\s^a_{\a\dot{\a}} \bar{\psi}^{\dot{\a}q}_a = 0,
\qquad
\psi^{~\a}_{aq} \s^a_{\a\dot{\a}} = 0,
\qquad
\end{equation}
can be obtained from selfconsistency conditions
of the equations \p{6.20}, \p{6.11'} and \p{6.12'} as described
 in \cite{bpstv,bsv} for $type~I$ super--p--branes and superstrings.

The selfconsistency conditions
also leads to  worldvolume supergravity torsion
constraints \cite{bsv,hs2}. In this respect it should be stressed that
choosing in \p{6.12'} the $m$ matrix in the form
\p{6.18} we get the main torsion constraint in the standard form
\begin{equation}\label{6.22}
T_{\a~\dot{\b}p}^{q~~c} = - i \d_{~p}^{q} \s^c_{\a\dot{\b}}.
\end{equation}

The  set of superfield equations for
super--$D$--$3$--brane \p{6.20}--\p{6.20'} obtained above generalizes
linearized equations for the D--3--brane studied in \cite{hs1} and is
similar to equations proposed for the $D=11$ super--$5$--brane in ref.
\cite{hs2}.

Thus the  generalized action proposed herein provides a bridge between the
formulations of refs. [1--4] and the superfield geometrical approach
of refs. \cite{bpstv,bsv,hs1,hs2}.

\section{Conclusion}

In conclusion we have proposed the Lorentz--harmonic formulation with
irreducible $\kappa$--symmetry and the generalized action functional for
Dirichlet super--p--branes in $D=10$ $type~II$ supergravity background.
In this formulation not only the WZ term but the whole super--D--p--brane
action is an integral of the  differential (p+1)--form. From the
generalized action we obtained the general form of the superfield
equations of motion for all super--D--p--branes and specified them in more
detail for $Type~IIB$ super--3--branes.

The superfield equations we obtained generalize linearized
super--D--p--brane equations of ref. \cite{hs1} and have the form
analogous to one proposed recently for $D=11$ super--5--branes in Ref.
\cite{hs2}.  Thus we have established a relation between the
``standard'' approach to super--D--p--branes based on the DBI action
\cite{c1}--\cite{bt} and the  superfield approach to these objects
\cite{hs1,hs2}.

A natural next step consists in studying the possibility of
constructing a generalized action for the  $D=11$ super--5--brane of
M--theory which should produce the superfield equations of ref. \cite{hs2}.

A covariant
action for the $D=11$ $5$--brane proposed recently in \cite{pst}
as a generalization of results of \cite{5} provides a basis for this
construction.

\bigskip
\noindent
{\bf Acknowledgments.}
The authors are grateful to K. Lechner and P. Pasti for useful
discussion.

Work of M.T. was supported by the European Commission TMR
programme \linebreak ERBFMRX--CT96--045 to which M.T. is associated.
I.B. and D.S. acknowledge partial support from the grant N2.3/644
of the Ministry of Science and Technology of Ukraine and
the INTAS Grants N 93--127, N 94--2317.

\end{document}